\documentclass[twocolumn,showpacs,amsmath,amssymb,10pt]{revtex4}
\usepackage{graphicx,color}
\usepackage{amsmath}
\usepackage{amssymb}

\usepackage{amsfonts}
\usepackage{bm}

\newcommand{\ot}{{\,\otimes\,}}
\newcommand{{\Cd}}{{\mathbb{C}^d}}

\newcommand{\tr}{\mathrm{Tr}}

\def\oper{{\mathchoice{\rm 1\mskip-4mu l}{\rm 1\mskip-4mu l}%
{\rm 1\mskip-4.5mu l}{\rm 1\mskip-5mu l}}}
\def\<{\langle}
\def\>{\rangle}

\begin{document}
\title{\textbf{On measures of non-Markovianity: divisibility vs. backflow of information}}
\author{Dariusz Chru\'sci\'nski$^1$, Andrzej Kossakowski$^1$ and \'Angel Rivas$^{2,3}$}
\affiliation{$^1$ Institute of Physics, Nicolaus Copernicus University \\
Grudzi\c{a}dzka 5/7, 87--100 Toru\'n, Poland\\
$^2$Departamento de F\'isica Te\'orica I, Facultad de Ciencias
F\'{\i}sicas, Universidad Complutense, 28040 Madrid, Spain\\
$^3$ Institut f\"{u}r Theoretische Physik,
Universit\"{a}t Ulm, Ulm D-89069, Germany}

\begin{abstract}
We analyze two recently proposed measures of non-Markovianity: one
based on the concept of divisibility of the dynamical map and the
other one based on distinguishability of quantum states. We provide
a toy model to show that these two measures need not agree. In addition,
we discuss possible generalizations and intricate relations between
these measures.
\end{abstract}


\maketitle

\section{Introduction}

The dynamics of open quantum systems attracts nowadays increasing
attention \cite{Breuer,Weiss,Alicki}.  It is  relevant not only for
the better understanding of quantum theory but it is fundamental in
various modern applications of quantum mechanics. Since the
system-environment interaction causes dissipation, decay and
decoherence it is clear that dynamic of open systems is fundamental
in modern quantum technologies, such as quantum communication,
cryptography and computation \cite{QIT}. The usual approach to the
dynamics of an open quantum system consists in applying an
appropriate Born-Markov approximation leading to the celebrated
quantum Markov semigroup \cite{GKS,Lindblad} which neglects all
memory effects. However, recent theoretical studies and
technological progress call for more refine approach based on
non-Markovian evolution.

Non-Markovian systems appear in many branches of physics, such as
quantum optics \cite{Breuer,Gardiner}, solid state physics
\cite{solid}, quantum chemistry \cite{Plenio-K}, and quantum
information processing \cite{Aharonov}. Since non-Markovian dynamics
modifies monotonic decay of quantum coherence it turns out that
when applied to composite systems it may protect quantum
entanglement for longer time than standard Markovian evolution
\cite{Saverio}. In particular it may protect the system against the
sudden death of entanglement \cite{DEATH}. It is therefore not
surprising that non-Markovian dynamics was intensively studied
during last years \cite{NM}.

Surprisingly, it turns out the concept of (non)Markovianity is not
uniquely defined. One approach is based on the idea of the
composition law which is essentially equivalent to the idea of
divisibility \cite{Wolf2}. This approach was used recently by Rivas,
Huelga and Plenio (RHP) \cite{RHP} to construct the corresponding
measure of non-Markovianity, that is, RHP measure the deviation from
divisibility. A different approach is advocated by Breuer, Laine and
Piilo (BLP) in Ref. \cite{BLP}. BLP define non-Markovian dynamics as
a time evolution for the open system characterized by a temporary
flow of information from the environment back into the system and
manifests itself as an increase in the distinguishability of pairs
of evolving quantum states. It is clear that RHP characterize a
mathematical property of the dynamical map whereas the idea of BLP
is based on physical features of  the system-reservoir interaction,
rather than  the mathematical properties of the dynamical map of the
open system. In a recent paper \cite{M} Haikka, Cresser and
Maniscalco performed detailed analysis of these approaches studying
the dynamics of the driven qubit in a structured environment. It is
indicated  \cite{M} that the concepts of RHP and BLP need not agree,
as was also conjectured in \cite{inne1} and checked in \cite{inne2}.
BLP measure was recently analyzed for the dynamics of a qubit
coupled to a spin environment via an energy-exchange mechanism
\cite{Apollaro}.

In the present paper we perform further analysis of this problem. In
particular we provide a simple toy model showing that Markovian
evolution \`a la BLP may be indivisible and hence non-Markovian
according to RHP. Actually, one may feel the relation BLP vs. RHP quite analog
to the relation between separable and PPT states in entanglement
theory, where separable states define the proper subset of PPT
states. States which are PPT but entangled are bound entangled.
Using this analogy one may call Markovian evolution \`a la BLP but
indivisible (non-Markovian according to RHP) -- `bound non-Markovian'.
Finally we discuss possible generalizations of BLP using tensor product
structures, and show that a slight modification of BLP reduces their concept of Markovianity to divisibility
(the bound non-Markovianity can be washed out). The main motivations of
this paper are to expose the relation between two different concepts
on a simple example, and to present the way to unify them,
which we hope clarify the issue.

\section{Divisibility vs. backflow of information}

The measure for non-Markovianity proposed by Rivas,
Huelga and Plenio (RHP) \cite{RHP} is based on a notion of divisibility:
a trace praserving completely positive map $\Lambda(t,0)$ is divisible if
it can be written as a
\begin{equation}\label{}
    \Lambda(t+\tau,0) = \Lambda(t+\tau,t)  \Lambda(t,0)\ ,
\end{equation}
and $\Lambda(t+\tau,t)$ is completely positive for any $t,\tau > 0$.
RHP define a map to be Markovian exactly when it is divisible.
Note that
\begin{equation}\label{}
    \Lambda(t+\tau,t) = \Lambda(t+\tau,0) \Lambda^{-1} (t,0)\ ,
\end{equation}
satisfies composition law
\begin{equation}\label{}
     \Lambda(s,t) = \Lambda(s,u) \Lambda(u,t)\ ,
\end{equation}
for any $s\geq u \geq t$, which is usually attributed for Markovian evolution.
It is shown \cite{RHP} that the quantity
\begin{equation}\label{g}
    g(t) = \lim_{\epsilon \rightarrow 0+} \frac{|| \oper_d \ot
    \Lambda(t+\epsilon,t)P^+_{d} ||_1 - 1}{\epsilon} \ ,
\end{equation}
enjoys $g(t) > 0$ if and only if the original map $\Lambda(t,0)$ is
indivisible ($||\cdot ||_1$ is a trace-norm). As usual $\oper_d$ denotes
an identity map in $M_d(\mathbb{C})$, and $P^+_d$ denotes the
maximally entangled state in $\mathbb{C}^d \ot \mathbb{C}^d$ (we
consider the evolution of a $d$-level system).  Now, using the fact
that any divisible (and differentiable) completely positive map satisfies a local in time
master equation
\begin{equation}\label{}
    \frac{d}{dt} \Lambda(t,0) = L(t) \Lambda(t,0)\ , \ \ \ \ \Lambda(0,0)=\oper_d\ ,
\end{equation}
where the local generator $L(t)$ is a legitimate Markovian generator for any $t\geq 0$.
The formula (\ref{g}) may be equivalently rewritten in terms of $L(t)$:
\begin{equation}\label{g-L}
    g(t) = \lim_{\epsilon \rightarrow 0+} \frac{|| [\oper_d \ot (\oper_d - \epsilon
    L(t))]P^+_{d} ||_1 - 1}{\epsilon} \ .
\end{equation}
A second criterion of non-Markovianity was proposed by Breuer, Laine
and Piilo (BLP) in Ref. \cite{BLP}. The BLP criterion identifies
non-Markovian dynamics with certain  physical features of the
system-reservoir interaction. They define non-Markovian dynamics as
a time evolution for the open system characterized by a temporary
flow of information from the environment back into the system. This
backflow of information may manifest itself as an increase in the
distinguishability of pairs of evolving quantum states. Hence,
according to BLP the dynamical map $\Lambda(t,0)$ is non-Markovian
if there exists a pair of initial states $\rho_1$ and $\rho_2$ such
that for some time $t>0$ the distinguishability of $\rho_1$ and
$\rho_2$ increases, that is,
\begin{equation}\label{}
    \sigma(\rho_1,\rho_2;t) = \frac{d}{dt}\, D[\rho_1(t),\rho_2(t)]\ ,
\end{equation}
where $D(\rho_1,\rho_2) = \frac 12 ||\rho_1-\rho_2||_1$ is the
distinguishability of $\rho_1$ and $\rho_2$, and $\rho_k(t) =
\Lambda(t,0)\rho_k$.

Using these two criteria one easily defines the corresponding non-Markovianity
measures of the dynamical map $\Lambda$:
\begin{equation}\label{NRHP}
    \mathcal{N}_{\rm RHP}(\Lambda) = \frac{\mathcal{I}}{\mathcal{I}+1} \ ,
\end{equation}
where
 $   \mathcal{I} = \int_0^\infty g(t)dt$.
Similarly, one has
\begin{equation}\label{NBLP}
    \mathcal{N}_{\rm BLP}(\Lambda) =
    \sup_{\rho_1,\rho_2}\, \int_{\sigma >0} \sigma(\rho_1,\rho_2;t)dt \ .
\end{equation}
It is clear \cite{BLP} that $\mathcal{N}_{\rm RHP}(\Lambda)=0$
implies $\mathcal{N}_{\rm BLP}(\Lambda)=0$. Hence, all divisible
maps are Markovian according to BLP. The converse is in general not
true. This problem was carefully analyzed in the paper \cite{inne2} for
phenomenological integro-differential master equations and in the recent paper \cite{M},
where the authors have studied the non-Markovian character of a
driven qubit in a structured reservoir in different dynamical regime
using $ \mathcal{N}_{\rm RHP}(\Lambda)$ and  $\mathcal{N}_{\rm
BLP}(\Lambda)$.

\section{Toy model -- classical stochastic dynamics}

In should be clear that the above discussion applies for a classical
stochastic systems as well. Consider stochastic dynamics of
$d$-state classical system described by the probability vector
$\mathbf{p}=(p_1,\ldots,p_d)$. Its evolution $\mathbf{p}(0)
\longrightarrow \mathbf{p}(t)=\Lambda(t) \cdot \mathbf{p}(0)$ is
defined by the family of stochastic matrices $\Lambda(t)$, that is,
$\Lambda_{ij}(t) \geq 0$ for all $i,j=1,\ldots,d$ and
\begin{equation}\label{}
    \sum_{i=1}^d \Lambda_{ij}(t) = 1\ ,
\end{equation}
for each $j=1,\ldots,d$ and $t \geq 0$. These conditions guarantee
that $\mathbf{p}(t)$ is a legitimate probability vector for all
$t\geq 0$. The corresponding local-in-time master equation
\begin{equation}\label{ME-c}
    \dot{p}_i(t) = \sum_{j=1}^d L_{ij}(t)\, p_j(t)\ ,
\end{equation}
is defined in terms of the local generator $L(t)$ which is defined
in by the classical dynamical map $\Lambda(t)$ as follows
\begin{equation}\label{G-class}
    L(t) = \dot{\Lambda}(t) \cdot \Lambda^{-1}(t)\ .
\end{equation}
In particular if $L$ does not depend on time $\Lambda(t)$ defines a
one-parameter semigroup of stochastic matrices $\Lambda(t) :=
\Lambda(t,0) = e^{tL}$. Let us recall that $\Lambda$ is stochastic if
if and only if it satisfies well known Kolmogorov conditions:
\begin{equation}\label{}
    L_{ij} \geq 0 \ , \ \ i \neq j \ , \ {\rm and}  \ \sum_{i=1}^d L_{ij}
    = 0 \ ,
\end{equation}
for each $j=1,\ldots,d$. Now, a stochastic evolution $\Lambda(t,0)$
is divisible if  it can be written as a composition of two
stochastic maps $\Lambda(t + \tau, t)$ and the original
$\Lambda(t,0)$
\begin{equation}\label{}
    \Lambda(t+\tau,0) = \Lambda(t+\tau,t) \cdot \Lambda(t,0)\ ,
\end{equation}
for any $\tau > 0$. Clearly, $\Lambda(t,0)$ is divisible if and only
if the corresponding local generator $L(t)$ satisfies Kolmogorov
conditions for any $t \geq 0$. Now, we would like to compare two
non-Markovianity measures due to RHP and BLP.

It is clear that classical evolution may be rewritten in the quantum
framework as follows: any probability vector $\mathbf{p}$ gives rise
to the diagonal density matrix $\rho = \sum_k p_k |k\>\<k|$, and
hence the stochastic map -- classical channel -- $\Lambda$ gives
rise to the following Kraus representation
\begin{equation}\label{}
    \Lambda\, \rho = \sum_{i,j=1}^d \Lambda_{ji}\, |i\>\<j|\, \rho\,
    |j\>\<i|\ .
\end{equation}
Now, to apply (\ref{g}) one needs the classical analog of $(\oper
\ot \Lambda)P^+$. Let us define $(\oper_{\rm cl} \ot
\Lambda)P^+_{\rm cl}$, where the classical identity map is defined
by
\begin{equation}\label{}
\oper_{\rm cl} \, \rho = \sum_{i=1}^d |i\>\<i|\, \rho\, |i\>\<i|\ ,
\end{equation}
and the classical analog of the maximally entangled states reads as
follows
\begin{equation}\label{}
P^+_{\rm cl} = \frac 1d \sum_{i=1}^d |i\>\<i|\ot |i\>\<i|\ .
\end{equation}
One finds therefore
\begin{equation}\label{}
(\oper_{\rm cl} \ot \Lambda)P^+_{\rm cl} = \frac 1d \sum_{i,j=1}^d
\Lambda_{ji}\, |i\>\<i|\ot |j\>\<j|\ .
\end{equation}
Now,  the  `classical' quantity
\begin{equation}\label{}
    g(t) = \lim_{\epsilon \rightarrow 0+} \frac{|| \oper_{\rm cl} \ot
    \Lambda(t+\epsilon,t)P^+_{\rm cl} ||_1 - 1}{\epsilon} \ ,
\end{equation}
is strictly positive if and only if the original stochastic map
$\Lambda(t,0)$ is indivisible and hence $g(t)$ identifies
non-Markovianity of the classical stochastic process. Equivalently,
it may be rewritten using local generator $L(t)$
\begin{equation}\label{}
    g(t) = \lim_{\epsilon \rightarrow 0} \frac{|| [\oper_{\rm cl} \ot (\oper_{\rm cl} \, -\,  \epsilon
    L(t))]P^+_{\rm cl} ||_1 - 1}{\epsilon} \ .
\end{equation}

Consider now the following toy model of stochastic dynamics of
2-state system
\begin{equation}\label{}
    \Lambda(t,0) = \left( \begin{array}{cc} 1 - x_0(t) & x_1(t) \\
    x_ 0(t) & 1- x_1(t) \end{array} \right) \ ,
\end{equation}
where $x_0(t), x_1(t) \in [0,1]$ for all $t \geq 0$. One finds for
the local generator
\begin{equation}\label{}
    L(t) = \left( \begin{array}{cc}  - a_0(t) & a_1(t) \\
    a_ 0(t) & - a_1(t) \end{array} \right) \ ,
\end{equation}
where
\begin{eqnarray} \label{a0}
  a_0 &=& \frac{\dot{x}_0(1-x_1) + \dot{x}_1x_0}{1-x_0-x_1} \ ,  \\
  \label{a1}
  a_1 &=& \frac{\dot{x}_0x_1 + \dot{x}_1(1-x_0)}{1-x_0-x_1} \ .
\end{eqnarray}
Now, $\Lambda(t,0)$ is divisible if and only if $a_0(t),a_1(t) \geq
0$ for all $t\geq 0$. On the other hand one easily finds
\begin{equation}\label{}
    \sigma(t) = -2[a_0(t) + a_1(t)]|\Delta_0|\ ,
\end{equation}
where $\Delta_k = (\mathbf{p}_1 - \mathbf{p}_2)_k$. Hence, contrary
to $g(t)$, the quantity $\sigma(t)$ controls only the sum of $a_0$
and $a_1$. In principle one may have $\sigma(t) \leq 0$ even if
$a_0(t) <0$ or $a_1(t) <0$. Indeed, let us define
\begin{equation}\label{}
    x_0(t) = \int_0^t f_0(\tau)d\tau\ , \ \ \ x_1(t) = \int_0^t f_1(\tau)d\tau\ ,
\end{equation}
such that $0 \leq \int_0^t f_k(\tau)d\tau \leq 1\,$, for all $t \geq
0$. Hence, our toy model is fully controlled by functions $f_0$ and
$f_1$. Let
\begin{equation}\label{}
    f_0(t) = \kappa \sin t\ , \ \ \ t\geq 0\ ,
\end{equation}
and $f_1(t) = 0$ for $t\in [0,\pi]$ together with
\begin{equation}\label{}
    f_1(t) = - \kappa \sin t\ , \ \ \ t\geq \pi\ ,
\end{equation}
where $0<\kappa<1/2$. One finds
\begin{equation}\label{}
    a_0(t) + a_1(t) = \frac{\kappa \sin t}{1-\kappa + \kappa \cos
    t}\ , \ \ \ t\in [0,\pi]\ ,
\end{equation}
and $a_0(t) + a_1(t) = 0$ for $t \geq \pi$. Note, that for $t \geq
\pi$ one has
\begin{equation}\label{}
    a_0(t) = - a_1(t) = \kappa \sin t\ ,
\end{equation}
which proves that $\Lambda(t,0)$ is not divisible. However, this map
is `classically' Markovian according to BLP \cite{BLP} due to $a_0(t) + a_1(t)
\geq 0$ for all $t\geq 0$.

\section{Qubit dynamics}

Consider now the following dynamics of a qubit
\begin{eqnarray}   \label{q-dyn}
  \rho_{00}(t) &=& \rho_{00}\, x_0(t) + \rho_{11}\, [ 1- x_1(t)] \ , \nonumber \\
  \rho_{11}(t) &=&  \rho_{00}\, [1-x_0(t)] + \rho_{11}\, x_1(t) \ ,\\
  \rho_{01}(t) &=& \rho_{01}\, \gamma(t) \ , \nonumber
\end{eqnarray}
where $x_0(t),x_1(t)\in [0,1]$, and
\begin{equation}\label{}
    |\gamma(t)|^2 \leq x_0(t) x_1(t)\ .
\end{equation}
The above conditions for $x_k(t)$ and $\gamma(t)$ guarantee that the
dynamics is completely positive. One easily finds for the
corresponding local generator
\begin{equation}\label{}
    L\, \rho = -i\frac{\Omega}{2} [\sigma_z,\rho] + \sum_{k=0}^1 a_k L_k \rho + \frac{\Gamma}{2} L_z \rho
    \ ,
\end{equation}
where
\begin{eqnarray}\label{L0L1Lz}
  {L}_0\rho &=& \sigma_+ \rho \sigma_- - \frac 12 \{ \sigma_-\sigma_+,\rho\} \ , \nonumber \\
  {L}_1\rho &=& \sigma_- \rho \sigma_+ - \frac 12 \{ \sigma_+\sigma_-,\rho\} \
  , \\
  {L}_z\rho &=& \sigma_z \rho \sigma_z - \rho\ . \nonumber
\end{eqnarray}
The time-dependent coefficients  $a_0$ and $a_1$ are defined in
(\ref{a0}) and (\ref{a1}), whereas $\Gamma(t)$ and $\Omega(t)$ read
as follows
\begin{eqnarray}
  \Gamma(t) &=& -\frac{a_0(t)+a_1(t)}{2}  - {\rm Re}\, \frac{\dot{\gamma}(t)}{\gamma(t)}  \ , \\
  \Omega(t) &=& {\rm Im} \, \frac{\dot{\gamma}(t)}{\gamma(t)}     \ .
\end{eqnarray}
The corresponding dynamical map $\Lambda(t,0)$ is divisible if and
only if $a_k(t), \Gamma(t) \geq 0$ for all $t \geq 0$. On the other
hand one finds the following formula
\begin{equation}\label{}
    \sigma(t) = - \frac{2A(t) \Delta_{00}^2 +
    [A(t) + 4\Gamma(t)]
    |\Delta_{01}|^2}{\sqrt{\Delta_{00}^2 + |\Delta_{01}|^2}}\ ,
\end{equation}
where $A(t) = a_0(t)+a_1(t)$, and a $2 \times 2$ matrix $\Delta$
reads
\begin{equation}\label{}
    \Delta = \rho_1(0) - \rho_2(0)\ .
\end{equation}
It is therefore  clear that using the same arguments as in the
classical toy model we may have $\sigma(t)\leq 0$ but the dynamical
map is not divisible.

\section{Generalized conditions}

We have studied so far the non-Markovian character of simple
classical and quantum (toy) models and compare two non-Markovianity
measures due to RHP and BLP. We performed explicit construction
showing that in general these two measures do not agree supporting
\cite{M,inne1,inne2}.

Let us observe that both approaches may be analyzed from a slightly
more general perspective. Consider a family $\Lambda(t,0)$ of trace
preserving positive maps (not necessarily completely positive).
Actually, the approach of BLP needs only positivity of
$\Lambda(t,0)$. It is clear that  $\Lambda(t,0)$ is contractive (one
has $||\Lambda(t,0)\rho||_1 = ||\rho||_1$). Now, $\Lambda(t,0)$ is
divisible if
\begin{equation}\label{}
    \Lambda(t+\tau,0) = \Lambda(t+\tau,t) \cdot \Lambda(t,0)\ ,
\end{equation}
and $\Lambda(t+\tau,t)$ is a contraction for all $\tau \geq 0$. It
implies
\begin{equation}\label{norm-1}
    ||\Lambda(t+\tau,t)||=1 \ ,
\end{equation}
where the norm of the map is defined via
\begin{equation}\label{}
    ||\Lambda|| = \sup_{||\rho||_1=1} ||\Lambda\rho||_1\ .
\end{equation}
Note, that $\Lambda(t,0)$ is divisible if and only if
\begin{equation}\label{}
    \frac{d}{dt}\Lambda(t,0) = L(t) \Lambda(t,0)\ ,
\end{equation}
where $L(t)$ is a local generator satisfying the following
dissipativity condition \cite{biuletyn} (see also \cite{GKS}):
for any set of mutually orthogonal rank-1 projectors $P_i$ such that $\sum_i P_i =
\mathbb{I}$, one has
\begin{eqnarray}
  {\rm Tr}[P_i L(t) P_j]  &\geq & 0\ , \ \ \ i\neq j \ , \nonumber \\
  \sum_i {\rm Tr}[P_i L(t) P_j] &=& 0\ .
\end{eqnarray}
The second condition guaranties that $\Lambda(t,0)$ is trace
preserving. Hence, formula (\ref{norm-1}) is equivalent to
\begin{equation}\label{norm-2}
 \lim_{\epsilon \rightarrow 0+} \frac{|| \oper - \epsilon L(t) || - 1}{\epsilon} =0 \
 .
\end{equation}
We call $\Lambda(t,0)$ completely divisible iff $\Lambda(t+\tau,t)$
is completely contractive. Recall  that $\Phi :
\mathcal{B}(\mathcal{H}) \longrightarrow  \mathcal{B}(\mathcal{H})$
is completely contractive \cite{Paulsen} iff $\oper_k \ot \Phi$ is
contractive for all $k=2,3,\ldots$. If ${\rm dim}\mathcal{H}= d <
\infty$, then, as in the case of complete positivity, it is enough
to check that $\oper_d \ot \Phi$ is contractive. Note, that due to
the fact that $\Lambda(t+\tau,t)$ is trace preserving complete
contractivity is equivalent to complete positivity. Hence, complete
divisibility is the standard divisibility considered by RHP
\cite{RHP}. Finally, recall  that $\Phi$ is completely bounded
\cite{Paulsen} iff $\oper_k \ot \Phi$ is bounded all $k=2,3,\ldots$.
One defines
\begin{equation}\label{}
    ||\Phi||_{(k)} = ||\oper_k \ot \Phi||\ ,
\end{equation}
and
\begin{equation}\label{}
    ||\Phi||_{\rm cb} = \sup_k ||\Phi||_{(k)}\ .
\end{equation}
$\Phi$ is completely contractive iff  $||\Phi||_{\rm cb} \leq 1$. If
${\rm dim}\mathcal{H}= d < \infty$, then necessarily $\Phi$ is
completely bounded. Moreover
\begin{equation}\label{}
    ||\Phi||_{\rm cb} = ||\Phi||_{(d)}\ .
\end{equation}
Hence, $\Lambda(t,0)$ is completely divisible iff
\begin{equation}\label{norm-3}
    ||\Lambda(t+\tau,t)||_{\rm cb} =1 \ ,
\end{equation}
which is much more restrictive than (\ref{norm-1}). Interestingly,
for completely positive $\Phi$ one has
\begin{equation}\label{}
    ||\Phi||_{\rm cb} = ||(\oper_d \ot \Phi)P^+_d||_1\ ,
\end{equation}
which reproduces result of RHP. Actually, one may generalize the
whole analysis to (completely) contractive dynamics on the Banach
space generalizing old results for contractive semigroups
\cite{Lumer}.

Now, let us observe that the approach of BLP which is valid for
positive maps $\Lambda(t,0)$ may be easily generalized if we assume
that $\Lambda(t,0)$ is completely positive. One may introduce
\begin{equation}\label{}
    \widetilde{\sigma}(\widetilde{\rho}_1,\widetilde{\rho}_2;t) = \frac{d}{dt}
    D[\widetilde{\rho}_1(t),\widetilde{\rho}_2(t)]\ ,
\end{equation}
where
\begin{equation}\label{}
    \widetilde{\rho}_k(t) = [\oper_d \ot
    \Lambda(t,0)]\widetilde{\rho}_k \ , \ \ k=1,2\ .
\end{equation}
It is clear that if $
\widetilde{\sigma}(\widetilde{\rho}_1,\widetilde{\rho}_2;t) \leq 0$
for all initial states $\widetilde{\rho}_k$, then
$\sigma(\rho_1,\rho_2;t) \leq 0$ for all initial states $\rho_k$.
Note, however, that the converse needs not be true. Again, one may
show that there exists indivisible maps $\Lambda(t,0)$ for which
$\widetilde{\sigma}(\widetilde{\rho}_1,\widetilde{\rho}_2;t) \leq 0$
for all initial states $\widetilde{\rho}_k$.
 Note, that  condition $
\widetilde{\sigma}(\widetilde{\rho}_1,\widetilde{\rho}_2;t) \leq 0$
 may be reformulated as
\begin{equation}\label{}
\widetilde{\sigma}(\Delta;t) = \frac{d}{dt}
    ||\Delta(t)||_1 \leq 0\ ,
\end{equation}
where $\Delta(t)= [\oper_d \ot \Lambda(t,0)]\Delta$, and
$\Delta^\dagger = \Delta$ together with ${\rm Tr}\Delta=0$, which
follows from $\Delta= (\widetilde{\rho}_1 - \widetilde{\rho}_2)/2$ and
$\widetilde{\rho}_k$ are true states. It should be stressed that condition
${\rm Tr}\Delta=0$ is very restrictive. Let us recall that if $\Phi$ is trace
preserving then $\Phi$ is positive iff
\begin{equation}\label{}
    ||\Phi\, a||_1 \leq ||a||_1\ ,
\end{equation}
for all $a^\dagger = a$. Note however that if one restricts to $a$
enjoying ${\rm Tr} a=0$, then $||\Phi\, a||_1 \leq ||a||_1$ does not
imply positivity of $\Phi$. If we relax ${\rm
Tr}\Delta=0$, then the condition $\widetilde{\sigma}(\Delta;t) \leq
0$ for all Hermitian $\Delta$ imply that $\Lambda(t,0)$ is
(completely) divisible and hence definition of Markovianity due to
BLP reduces to divisibility. More explicitly this fact can be formulated as:
\bigskip

\noindent \textbf{Theorem}. {\it For a bijective evolution
$\Lambda(t,0)$, $\tilde{\sigma}(\Delta;t)\leq0$ for all
$\Delta^\dagger=\Delta$, if and only if it is divisible.}
\bigskip

\noindent \textit{Proof}. The `if' part is straightforward because
the trace norm is monotonically decreasing given the completely
contracting property of the $\Lambda(t+\tau,t)$,
\[
\|\Lambda(t+\tau)\|_1=\|[\oper_d \ot
\Lambda(t+\tau,t)]\Delta(t)\|_1\leq\|\Delta(t)\|_1\ ,
\]
for every $t$ and $\tau>0$. Conversely, if the evolution
$\Lambda(t,0)$ is bijective the partitions
$\Lambda(t+\tau,t)=\Lambda(t+\tau,0)\Lambda^{-1}(t,0)$ are
well-defined. Then $\tilde{\sigma}(\Delta;t)>0$ for some $t$ implies
that for some small $\tau$
\[
\|\Delta(t+\tau)\|_1=\|[\oper_d \ot
\Lambda(t+\tau,t)]\Delta(t)\|_1>\|\Delta(t)\|_1.
\]

So there exists some partition $\Lambda(t+\tau,t)$ which is not
completely contractive, and therefore the dynamics is not divisible.
\begin{flushright}$\Box$
\end{flushright}

\section{Operational meaning of $\tilde{\sigma}(\Delta;t)$}

One may think that by relaxing the condition $\mathrm{Tr}\Delta=0$ the interpretation of an increase of $\tilde{\sigma}(\Delta;t)$ in terms of backflow of information from environment to system is lost, however that is not the case. To explain this more carefully we have to recall here some results from quantum hypothesis testing, particularly the one-shot two-state discrimination problem \cite{Hayashi}.

Consider a quantum system whose state is represented by the density matrix $\rho_1$ with probability $p$, and $\rho_2$ with probability $(1-p)$. We want to determine the density matrix that describes the true state of the quantum system by performing a measurement. If we consider some general POVM $\{M_j\}$, where $j\in\Omega$ is the set of possible outcomes, we may split this set in two cases. If the outcome of the measurement is inside of some subset $A\subset\Omega$, then we say that the state is $\rho_1$. Conversely if the result of the measurement belongs to the complementary set $A^c$ such that $A\cup A^c=\Omega$, we say that the state is $\rho_2$. Let us group the results of this measurement in another POVM given by the operator $T=\sum_{j\in A} M_j$.

Thus, when the true state is $\rho_1$ (which happens with probability $p$) we erroneously conclude that the state is $\rho_2$ with probability
\begin{eqnarray*}
(1-p)\sum_{j\in A^c} \tr[\rho_1 M_j]&=&(1-p)\tr\left[\rho_1 \left(\sum_{j\in A^c} M_j\right)\right]\\
&=&(1-p)\tr\left[\rho_1 (\mathbb{I}-T)\right].
\end{eqnarray*}
On the other hand, when the true state is $\rho_2$, we erroneously conclude that the state is $\rho_1$ with probability
\[
p\sum_{j\in A} \tr[\rho_2 M_j]=p\tr\left[\rho_2 \left(\sum_{j\in A} M_j\right)\right]=p\tr\left[\rho_2 T\right].
\]
Note that when $p=0$ or $p=1$ we immediately obtain zero probability to identify wrongly the true state. The problem in one-shot two-state discrimination is to examine the tradeoff between the two error probabilities $p\tr\left[\rho_2 T\right]$ and $(1-p)\tr\left[\rho_1 (\mathbb{I}-T)\right]$. Thus, consider the best choice of $T$ that minimizes the total error probability
\begin{eqnarray*}
\min_{0\leq T\leq \mathbb{I}} \{p\tr\left[\rho_2 T\right]+(1-p)\tr\left[\rho_1 (\mathbb{I}-T)\right]\}\\
=\min_{0\leq T\leq \mathbb{I}} \{(1-p)+\tr\left[p\rho_2 T-(1-p)\rho_1 T\right]\}\\
=(1-p)-\max_{0\leq T\leq \mathbb{I}} [\tr\left(\Delta T\right)],
\end{eqnarray*}
where $\Delta=(1-p)\rho_1-p\rho_2$ is a self-adjoint operator (sometimes called Helstrom matrix \cite{Helstrom}) with trace $\tr \Delta=1-2p$, which vanishes only for the unbiased case $p=1/2$. By using analogous arguments that for $p=1/2$ (see \cite{QIT,Hayashi}) the result of the optimization process turns out to be
\begin{equation}
\min_{0\leq T\leq \mathbb{I}} \{p\tr\left[\rho_2 T\right]+(1-p)\tr\left[\rho_1 (\mathbb{I}-T)\right]\}=\frac{1-\|\Delta\|_1}{2}.
\end{equation}

Thus the trace norm of $\Delta=(1-p)\rho_1-p\rho_2$ gives our capability to distinguish correctly between $\rho_1$ and $\rho_2$ in the one-shot two-state discrimination problem. Since that is only a function of the information we gather by the prior probability $p$ and by the measurement $T$, $\|\Delta\|_1$ is a measure of the information we have.

Consider again a trace preserving map $\Phi
:\mathcal{B}(\mathcal{H}) \longrightarrow \mathcal{B}(\mathcal{H})$.
If it increases the trace norm of $\Delta$,
$\|\Phi(\Delta)\|_1>\|\Delta\|_1$, we can assume that $\Phi$ carries
information about the correct state of the system. Otherwise it
cannot decrease the probability to identify wrongly the true state!.
For that reason, if we process data before making a measurement,
which means to apply some $\Phi$ over the states, we cannot increase
the trace norm of $\Delta$; since we cannot gain more information
about some data just by processing it!. This is the so-called data
processing inequality \cite{Hayashi}, and implies that $\Phi$ is
contractive, i.e. positive. In addition, we can never dismiss that
the states $\rho_1$ and $\rho_2$ are part of a larger system ``SA''.
In such a way that they are the result of tracing out the ancillary
systems $\rho_{1,2}=\mathrm{Tr}_A[\rho_{1A,2A}]$. By making the same
analysis as before, in order not to gain information we then require
the map $\oper_d \ot \Phi$ to be contractive, i.e. $\Phi$ completely
contractive and thus completely positive.

Going back to the problem of Markovianity, one may now realize that $\tilde{\sigma}(\Delta;t)$ is a measure of the information gained by the system for some initial Helstrom matrix $\Delta$. This interpretation puts together the two concepts of Markovianity based upon divisibility and backflows of information. Particularly if one ignores the potential existence of an ancilla, under the case of unbiased discrimination problem, $\|\Delta\|_1=\frac{1}{2}\|\rho_1-\rho_2\|_1$, which is the definition of the trace distance between $\rho_1$ and $\rho_2$, recovering the approach of BLP. Nevertheless in principle there is no reason to presuppose the fulfillment of these particular assumptions.

One can define a measure of non-Markovianity as (\ref{NBLP}), with $\tilde{\sigma}(\Delta;t)$ in the place of $\sigma(\rho_1,\rho_2;t)$. That will be zero iff (\ref{NRHP}) is, however (\ref{NRHP}) is easier to compute because it avoids the complicated optimization procedure in (\ref{NBLP}) now upon every Helstrom matrix $\Delta$.

We have already proposed examples of non-divisible dynamics with
$\mathcal{N}_{BLP}(\Lambda)=0$, but for completeness we have
computed $\tilde{\sigma}(\Delta;t)$ for the phenomenological
integro-differential master equation \cite{pheno}:
\begin{equation}\label{int-diff}
\frac{d\rho(t)}{dt}=\int_0^tk(t')L\rho(t-t')dt',
\end{equation}
where $k(t)=\gamma e^{-\gamma t}$ and
\begin{equation}\label{}
L\rho=\gamma_0(\bar{n}+1)L_0\rho+\gamma_0\bar{n}L_1\rho,
\end{equation}
with $\gamma,\gamma_0,\bar{n}\geq0$ and $L_0$ and $L_1$ were given
in (\ref{L0L1Lz}). Despite this kind of equations is used as a
simple model of non-Markovian dynamics, it was showed in
\cite{inne2} that $\mathcal{N}_{BLP}(\Lambda)=0$. However
$\tilde{\sigma}(\Delta;t)$ is not always decreasing for any Helstrom
matrix. For example by taking $p=0.07$ and
\begin{eqnarray}\label{rho1}
\rho_1&=&\left(\begin{array}{cccc}
0.5& 0& 0& 0.48\\
0& 0.001& 0& 0\\
0& 0& 0.019& 0\\
0.48& 0& 0& 0.48
\end{array}\right),\\ \label{rho2}
\rho_2&=&\left(\begin{array}{cccc}
0.25& 0& 0& 0\\
0& 0.25& 0& 0\\
0& 0& 0.5& 0\\
0& 0& 0& 0
\end{array}\right),
\end{eqnarray}
there is a period where $\tilde{\sigma}(\Delta;t)$ grows as we have
represented in Fig 1. This denotes the existence of a backflow of
information which was expected from the non-divisible character of
this dynamics.
\begin{figure}
\begin{center}
\includegraphics[width=0.5\textwidth]{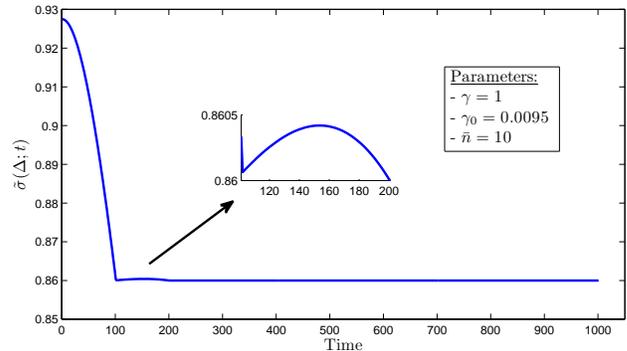}
\end{center}
\caption{Behavior of $\tilde{\sigma}(\Delta;t)$ for the
integro-differential model (\ref{int-diff}). The initial Helstrom
matrix is $\Delta=0.93\rho_1-0.07\rho_2$, where $\rho_1$ and
$\rho_2$ are given in (\ref{rho1}) and (\ref{rho2}). There is a time
period between 100 and 150 where $\tilde{\sigma}(\Delta;t)$
increases.}
\end{figure}

\section{Conclusions}
We have analyzed two concepts of Markovianity, one based on the
divisibility property of the dynamical map and the other based upon
the distinguishability of quantum states. We have given very simple
examples where these two criteria do not coincide. Furthermore we
have proposed a way to make them equivalent, in the sense that
Markovianity would be identified by divisibility, but keeping the
interpretation in terms of flows of information. For that we resort
to the results in the one-shot two-state discrimination problem; and
point out that an increase of the trace norm of any Hermitian matrix
during the dynamics can also be associated with a backflow of
information from the environment to the system.

\acknowledgements
A.R. is grateful to Michael M. Wolf, Susana F. Huelga and Martin B. Plenio for enlightening discussions about this topic, and thanks financial support from the EU Integrated Project QESSENCE, project QUITEMAD S2009-ESP-1594 of the Consejer\'{\i}a de Educaci\'{o}n de la Comunidad de Madrid and MICINN FIS2009-10061.

\end{document}